\documentclass[twocolumn]{aastex63}

\usepackage{apjfonts}
\usepackage{color}

\usepackage{graphicx}

\shorttitle{Infrared observations of V1047~Cen}
\shortauthors{T. R. Geballe et al.}

\begin{document}
\title{Infrared spectroscopy of the recent outburst in V1047 Cen (Nova Centauri 2005)}

\author[0000-0003-2824-3875]{T. R. Geballe}
\affiliation{Gemini Observatory, 670 N. A'ohoku Place Hilo, Hawaii, 96720, USA}

\author{D. P. K. Banerjee}
\affiliation{Physical Research Laboratory, Navrangpura, Ahmedabad, Gujarat 380009, India}

\author[0000-0002-3142-8953]{A. Evans}
\affiliation{Astrophysics Group, Keele University, Keele, Staffordshire,
ST5 5BG, UK}

\author[0000-0003-1319-4089]{R. D. Gehrz}
\affiliation{Minnesota Institute for Astrophysics, School of Physics \& Astronomy,
116 Church Street SE, University of Minnesota, Minneapolis, MN 55455, USA}

\author[0000-0001-6567-627X]{C. E. Woodward}
\affiliation{Minnesota Institute for Astrophysics, School of Physics \& Astronomy,
116 Church Street SE, University of Minnesota, Minneapolis, MN 55455, USA}

\author[0000-0001-7016-1692]{P. Mr\'oz}
\affiliation{Astronomical Observatory, University of Warsaw, Al. Ujazdowskie 4, 
00-478 Warsaw, Poland}

\author[0000-0001-5207-5619]{A.Udalski}
\affiliation{Astronomical Observatory, University of Warsaw, Al. Ujazdowskie 4, 
00-478 Warsaw, Poland}

\author[0000-0001-6805-9664]{U. Munari}
\affiliation{INAF Astronomical Observatory of Padova, I-36012 Asiago (VI), Italy}

\author[0000-0002-1359-6312]{S. Starrfield}
\affiliation{School of Earth and Space Exploration, Arizona State University, 
Box 871404, Tempe, AZ 85287-1404, USA}

\author[0000-0001-5624-2613]{K. L. Page}
\affiliation{School of Physics and Astronomy, University of Leicester, University Road, Leicester, LE1 7RH, UK}

\author[0000-0001-5991-6863]{K. Sokolovsky}
\affiliation{Department of Physics and Astronomy, Michigan State University,
567 Wilson Rd, East Lansing, MI 48824, USA}
\affiliation{Sternberg Astronomical Institute, Moscow State University,
Universitetskii~pr.~13, 119992~Moscow, Russia}
\affiliation{Astro Space Center of Lebedev Physical Institute,
Profsoyuznaya~St.~84/32, 117997~Moscow, Russia}

\author{F.-J. Hambsch}
\affiliation{Vereniging Voor Sterrenkunde (VVS), Oostmeers 122 C, 8000 
Brugge, Belgium}
\affiliation{Bundesdeutsche Arbeitsgemeinschaft f\"ur Ver\"anderliche Sterne e.V. (BAV), Munsterdamm 90, D-12169 Berlin, Germany}
\affiliation{American Association of Variable Star Observers, 49 Bay State
Road, Cambridge, MA 02138, USA}

\author[0000-0002-9810-0506]{G. Myers}
\affiliation{American Association of Variable Star Observers, 49 Bay State
Road, Cambridge, MA 02138, USA}
\affiliation{5 Inverness Way, Hillsborough, CA 94010}

\author[0000-0001-8525-3442]{E. Aydi}
\affiliation{Center for Data Intensive and Time Domain Astronomy, Department of Physics and Astronomy, Michigan State University, East Lansing, MI 48824, USA}

\author[0000-0002-7004-9956]{D. A. H. Buckley}
\affiliation{South African Astronomical Observatory, P.O. Box 9,
Observatory 7935, South Africa}

\author[0000-0001-7796-1756]{F. Walter}
\affiliation{Department of Physics and Astronomy, Stony Brook University, 
Stony Brook NY 11794-3800, USA}

\author[0000-0003-1892-2751]{R. M. Wagner}
\affiliation{Department of Astronomy, The Ohio State University, 
140 W. 18th Avenue, Columbus, OH 43210, USA}
\affiliation{Large Binocular Telescope Observatory, 933 North Cherry Avenue,
Tucson, AZ 85721, USA}

\newcommand{\vdag}{(v)^\dagger}
\newcommand\aastex{AAS\TeX}
\newcommand\latex{La\TeX}

\newcommand{\Mdot}[2]{\mbox{${#1}\times10^{-{#2}}$\,M$_\odot$~yr$^{-1}$}}
\newcommand{\Msun}{\mbox{\,M$_\odot$}}
\newcommand{\Lsun}{\mbox{\,L$_\odot$}}
\newcommand{\vunit}{\mbox{\,km\,s$^{-1}$}}
\newcommand{\mic}{\mbox{$\,\mu$m}}
\newcommand{\pion}[2]{{#1}\,{\sc {#2}}}
\newcommand{\fion}[2]{[{#1}\,{\sc {#2}}]}
\newcommand{\Ne}{\mbox{$n_{\rm e}$}}
\newcommand{\Te}{\mbox{$T_{\rm e}$}}
\newcommand{\ltsimeq}{\raisebox{-0.6ex}{$\,\stackrel
        {\raisebox{-.2ex}{$\textstyle <$}}{\sim}\,$}}
\newcommand{\gtsimeq}{\raisebox{-0.6ex}{$\,\stackrel
        {\raisebox{-.2ex}{$\textstyle >$}}{\sim}\,$}}
\newcommand{\prsimeq}{\raisebox{-0.6ex}{$\,\stackrel
        {\raisebox{-.2ex}{$\propto$}}{\sim}\,$}}

\newcommand{\vcen}{\mbox{V1047~Cen}}

\newcommand{\ncrit}{\mbox{$n_{\rm crit}$}}

\newcommand{\chemone}{\raisebox{0.03cm}{$-$}} 
\newcommand{\chemtwo}{\raisebox{0.03cm}{$=$}} 
\newcommand{\chemthree}{\raisebox{0.03cm}{$\equiv$}} 
\newcommand{\chemthreehalf}{$-$\hspace{-0.2cm}\raisebox{0.18cm}
           {\footnotesize ...}} 

  \DeclareGraphicsExtensions{.png,.pdf,.eps}

\begin{abstract}

Fourteen years after its eruption as a classical nova (CN),
\vcen\ (Nova Cen 2005) began an unusual re-brightening in 2019 April.
The amplitude of the brightening suggests that this is a dwarf nova (DN) eruption
in a CN system. Very few CNe have had DN eruptions within decades of the main CN 
outburst. The 14 years separating the CN and  DN eruptions of \vcen\
is the shortest of all instances recorded thus far.
Explaining this rapid succession of CN and DN outbursts
in \vcen\ may be challenging within the framework of standard theories for
DN outbursts. Following a CN eruption, the mass
accretion rate is believed to remain high $(\dot{M}\sim10^{-8}$\Msun~yr$^{-1})$
for a few centuries, due to the
irradiation of the secondary star by the still-hot surface of the white dwarf.
Thus a DN eruption is not expected to occur during this high mass
accretion phase as DN outbursts, which result from thermal instabilities
in the accretion disk, and arise during a regime of low mass
accretion rate $(\dot{M}\sim10^{-10}$\Msun~yr$^{-1})$. Here we present
near-infrared spectroscopy to
show that the present outburst is most likely a DN eruption, and discuss the
possible reasons for its early occurrence. Even if the present re-brightening
is later shown to be due to a  cause other than a DN outburst,
the present study provides invaluable documentation of this unusual event.

\end{abstract}


\section{Introduction}
\label{intro}

A Cataclysmic Variable (CV) generally consists of a compact star
(the primary) in a semi-detached binary system with a donor star that
transfers matter onto the primary through the inner Lagrangian point
via an accretion disk (AD). In classical and dwarf nova (CN and DN
respectively) systems the primary is a white dwarf (WD), and the
secondary is a late-type star \citep[see][]{warner}.

A CN eruption is the result of a thermonuclear runaway (TNR) in the 
degenerate 
material accreted on the surface of the WD \citep[see][]{CN2}. Following the
TNR, $\sim10^{-6}-10^{-4}$\Msun\ of material, enriched in CNO, is ejected
explosively at several 100--1000\vunit. The system luminosity increases
by a factor of as much as $\sim15$~mag. The bolometric luminosity remains
approximately constant during the eruption, so the visual flux declines
as the effective temperature of the stellar remnant increases. Once 
the ejecta become optically thin to 0.1--1.0~keV photons,
the source becomes a ``super-soft X-ray source''
\citep[see][]{krautter}. 
CN eruptions are believed to repeat on time-scales of $\sim10^4$~years.

Eruptions of DNe of the U~Gem type are gentler affairs, and are due 
to an increase of the mass-flow through the AD as a consequence of a
thermal instability within
the AD itself; the  amplitude of the outburst, which typically lasts
a few days, is 
{generally $\ltsimeq5$~mag,
although WZ~Sge stars can have outburst 
amplitudes up to 8~mag \citep{warner}}. 
There is little or no material ejected,
and there is no enhancement of elemental abundances.

A small number of CNe have, since their eruptions, shown DN outbursts
\citep[][and references therein]{livio,mroz}, while some DNe seem
to have circumstellar shells that appear to be the result of a
previous CN outburst \citep[e.g.][]{shara-nat}. 
In this cyclical scheme, CN--DN--CN
\citep{shara-apj86}, a DN eruption is not expected to
follow within a decade of the CN outburst. This is because
the cooling time-scale for the WD after the CN eruption is of the order
of a century or more, during which high mass transfer rates 
from the secondary continue due to irradiation by the WD
\citep{shara-apj86,kovetz}. Such circumstances are not conducive for a
DN eruption unless special circumstances prevail (see
Section~\ref{disc} below).  Hence
the present outburst in V1047 Cen, if it is indeed a DN eruption,
is a rare and exceptional event.

Here we present two epochs of  near-infrared (NIR) spectroscopy of the
CN \vcen\ taken around the peak brightness of its
2019 outburst, and also present optical and NIR light curves that document
the detection and evolution of the present re-brightening event.

\section{\vcen\ (Nova Centauri 2005)}

\begin{figure}
\includegraphics[width=7.5cm,keepaspectratio]{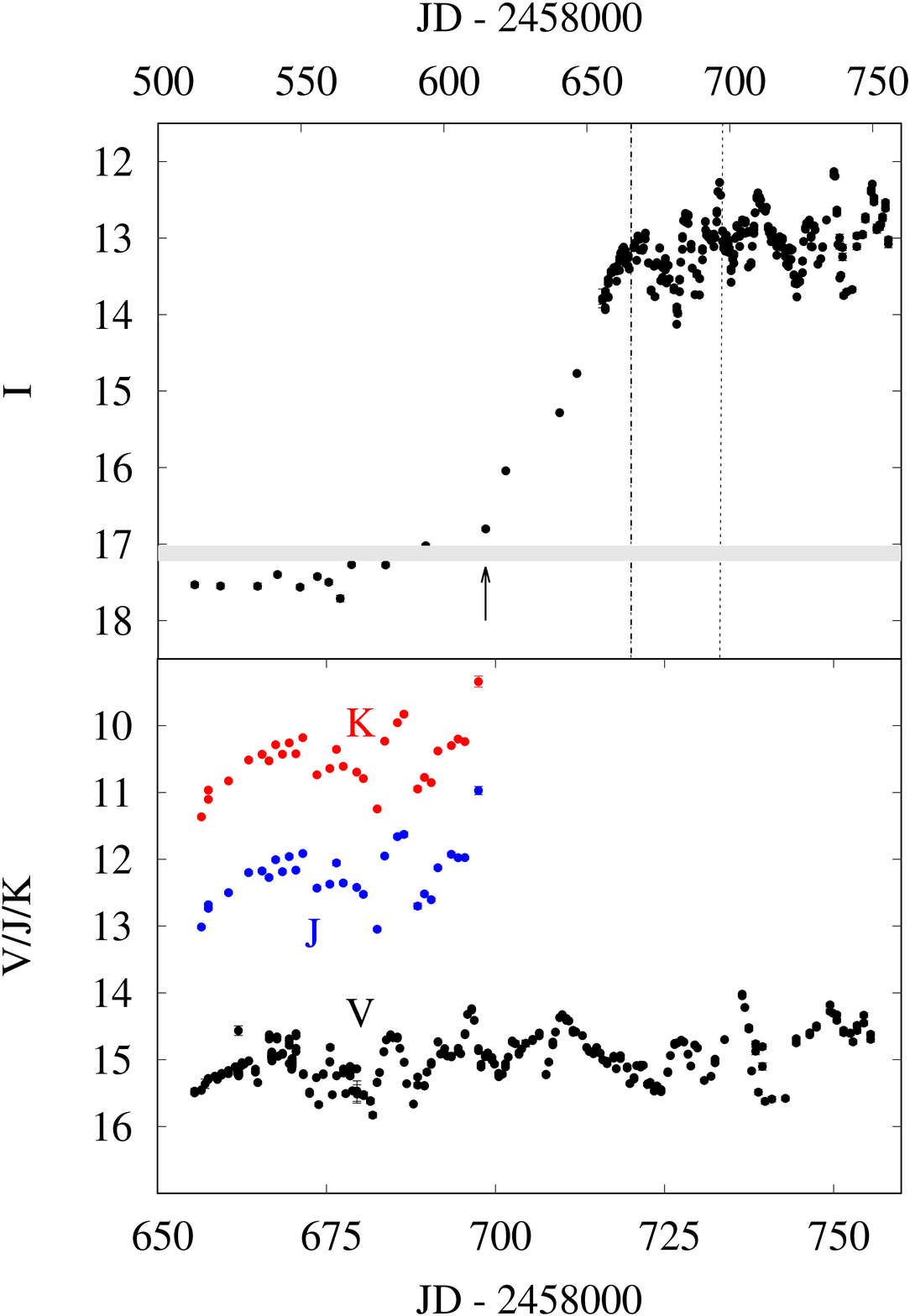}
\caption{Top: $I$ band light curve of \vcen, showing the 2019 April outburst;
data from the OGLE survey. Approximate time of the eruption is
shown by the arrow. Horizontal shading denotes mean
pre-2019 outburst $I$ magnitude
and $\pm1\sigma$ variation from \cite{mroz2}.
Times of the Gemini observations are indicated by the vertical dotted
lines. Bottom: $V\!JK$ light curve of \vcen\ during eruption; note
different JD scales.
Data from the AAVSO and SMARTS \citep{walter} databases.\label{LC}}
\end{figure}

\vcen\ (Nova Cen 2005) was discovered by \cite{liller} on 2005 
September 1.031 at $m_{\rm vis}\simeq8.5$, but the 2005 eruption 
was not well observed.
It was a CN of the ``\pion{Fe}{ii}'' spectroscopic
type \citep{walter}. The H$\alpha$ line was found to have 
a FWHM of $\sim840$\vunit\ on day~3 \citep{liller}, 
and $\sim1200$\vunit\ on day~6.9 \citep{walter}.

\vcen\ (AT2019hik, Gaia19cfn) was found to be in outburst again by
the Optical Gravitational Lensing Experiment \citep[OGLE;][]{ogle}
survey \citep[][see Fig.~\ref{LC}]{mroz2}. 
The outburst started on 2019 April 6 and, at the time of writing
(2019 mid-October, when it approached solar conjunction), had persisted
for $\sim160$~days. The nature of the slow rise to maximum, 
{the outburst amplitude,}
and the subsequent variability, 
rule out a recurrent nova eruption for the 2019 event 
in \vcen\ \citep[see][]{RN}. Currently, the  duration of the 2019 April outburst
is of the same order as that of DN eruptions in the 
long period systems, and considerably
longer than those in GK~Per \citep[e.g.][]{szkody, salazar}. The 
likelihood is that we are seeing a protracted DN eruption in a CN system.

High-resolution optical spectroscopy shortly after the 2019 April outburst was
reported by \cite{aydib}. At that time the spectrum was dominated by
\pion{H}{i} ($\mbox{FWZI}>2500$\vunit) and \fion{O}{iii}
($\mbox{FWZI}>1100$\vunit), which they suggest might have arisen in
material ejected in the 2005 eruption. There were also narrow \pion{He}{i}
emission lines ($\mbox{FWZI}<300$\vunit), which may have originated in
the 2019 DN eruption.

Timing analysis of the \vcen\ light curve from the Transiting Exoplanet
Survey Satellite \citep[TESS;][]{tess} indicated a possible $8.66\pm0.07$~hour 
periodicity \citep{aydia}, although this has yet to be verified. If 
this is confirmed to be the orbital period,
it is at the higher end of CV orbital periods, and possibly suggests
an evolved secondary.

The  distance and reddening to \vcen\ are poorly constrained. 
The {\it Gaia} survey Data Release 2 \citep[][see this paper for the
distance derivation, and its reliability]{gaia}
gives an estimated distance of 1.86~kpc, with lower and upper confidence
bounds of 1.01~kpc and 4.25~kpc respectively.
This renders it difficult to assign an accurate value for the  reddening 
using $3D-$extinction maps. For example, the extinction in the direction 
of \vcen\ from  \cite{marshall} gives $E(B-V)$ values of 0.29, 0.54 and 
2.54 at the {\it Gaia} lower, most probable, and upper distance estimates. Literature values for $E(B-V)$ give a 
wide range  \citep[e.g][give $E(B-V)$ in the range 
1.28--1.38]{senziani}.
Using the \pion{He}{ii} $\lambda4686$\AA/$\lambda$10124\AA\ line ratio,
$E(B-V)=1.1$ is suggested by \cite{dimille}, who also used the 5780\AA\
diffuse interstellar band to estimate $E(B-V)=1$.

In view of the considerable uncertainty in
$E(B-V)$ we do not deredden the data here.
 
\section{Observations and Data Reduction}

\begin{table}
\caption{Log of spectroscopic observations.\label{log}}
\centering
\begin{tabular}{l c r c c c c } \hline
Date &   Grism, coverage   &  IT   & Standard star \\
    &   resolution\footnote{With the 0\farcs36 slit, for the $J\!H$  grism,  
    $R$ ranges from $\sim200$ near the short wavelength edge, to a maximum of
$\sim1100$ at 1.3--1.4\mic, dropping to $\sim400$ near the long
wavelength edge. For the R3K grism, $R$ varies from 1200 near 1.9\mic,
increasing  to 3200 near 2.2\mic, and then decreasing  to $\sim1200$
at 2.4\mic.}   &  (s)  &  \\ \hline
2019 Jun 30/Jul 1 &  R3K, 1.9--2.5\mic   & 1920 & HIP 63036& \\
2019 Jun 30/Jul 1 &  JH, 0.9--1.8\mic    & 240 & HIP 67360& \\
2019 Aug 1/2      &  R3K, 1.25--2.45\mic & 1260 & HIP 67360& \\
2019 Aug 1/2      &  JH, 0.9--1.8\mic    & 240 & HIP 67360& \\ \hline
\end{tabular}
\end{table}

\begin{figure*}
\includegraphics[width=13cm,keepaspectratio]{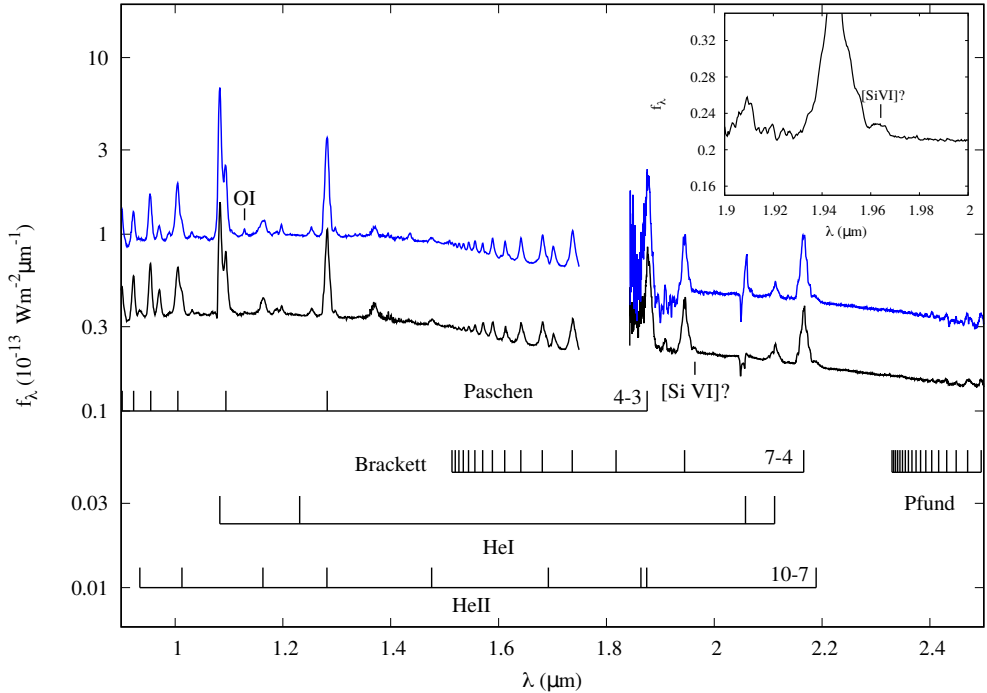}
\caption{0.9--2.5\mic\ spectra of \vcen\ obtained on 2019 June 30/July 1
(black) and 2019 August 1/2 (blue). Hydrogen recombination and other
lines are identified; for recombination lines, the transition for the 
longest wavelength member is shown.
Inset shows the putative \fion{Si}{vi} coronal line.
The data behind this figure will be available in the online version of the paper after publication.
\label{gemini}}
\end{figure*}

\setcounter{footnote}{0}
Spectra of \vcen\ were obtained at the 8.1~m Gemini South Telescope using its
facility spectrograph FLAMINGOS-2 \citep{eik04} in conditions
of poor seeing and possible thin clouds. The log of the observations
is given in Table~\ref{log}.
Data reduction was done using standard procedures for NIR data 
using IRAF\footnote{IRAF is distributed by the National Optical Astronomy
Observatories, which is operated by the Association of Universities for
Research in Astronomy, Inc. (AURA) under cooperative agreement with the National
Science Foundation.}
and Figaro tasks. Because of the poor conditions under which the
spectra were obtained, flux calibration of the spectra used broad
band NIR photometry from SMARTS (see Fig.~\ref{LC}). The uncertainty
in the flux scaling is 20\% for the June~30/July~1 spectrum, 
and higher for the August~1/2 spectrum, as no NIR photometry was obtained on that night. However as we do not use absolute flux values 
in this study, our results and conclusions are not affected.
The observed spectra are shown in Fig.~\ref{gemini}.

\section{The infrared spectra}

\begin{figure}
\includegraphics[width=6.cm,keepaspectratio]{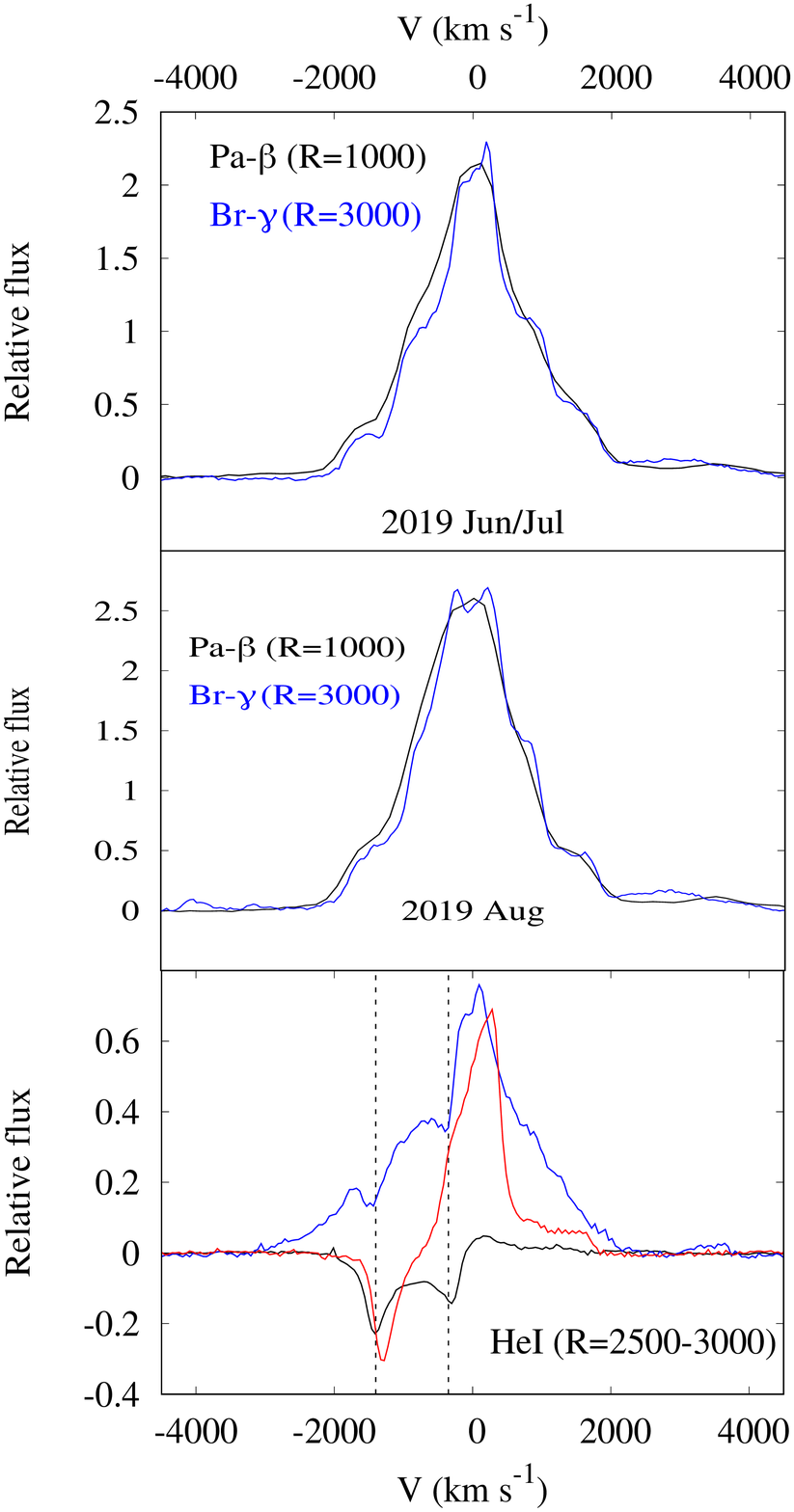}
\caption{Top and middle: profiles of Pa-$\beta$ and Br-$\gamma$ lines, 
expressed as a function of velocity.
Bottom: profiles of the \pion{He}{i} 2.0587\mic\ singlet for 2019
Jun/Jul (black) and Aug (red), and \pion{He}{i} 2.1138\mic\ for Jun/Jul
(blue).
{Vertical lines denote troughs discussed in text.}
Numbers in brackets are spectral resolutions.
\label{vel}}
\end{figure}

{The IR spectra rule out the possibility that \vcen\ is 
a stellar merger event (or ``Luminous Red Nova''),
as suggested by \cite{mroz2}. While stellar mergers have shown 
a similar slow rise to maximum, they also show a distinct shift to 
cooler spectral type, and display very specific NIR molecular 
features (e.g. AlO, H$_2$O) within 45 days of eruption.
Our data show none of these characteristics.}
{The absence of first overtone CO bands in the IR
spectra (see Fig.~\ref{gemini}) also rules out the possibility that
\vcen\ might be a symbiotic system.}

Both spectra show lines of \pion{H}{i}, \pion{He}{i} and \pion{He}{ii}.
The C, N and O lines that are generally seen in the NIR spectra of RNe -- irrespective of whether 
the RN is of the T Pyx, T CrB or U Sco sub-types 
\citep{banerjee09,banerjee10,joshi}
-- are absent in \vcen, with the exception of the \pion{O}{i} 
line at 1.1289\mic\ in August.

We have determined the velocities from the FWZIs of the \pion{H}i
Pa-$\beta$ and Br-$\gamma$ lines (Fig.~\ref{vel}) to be $\sim2000$\vunit, 
as reported by \cite{aydib} for the optical \pion{H}{i} lines, with 
no significant change between the two IR observations. The FWHMs of 
these lines are $\sim1,500$\vunit, somewhat larger than the 
velocities reported by \cite{liller} and \cite{walter} for H-$\alpha$ 
during the 2005 eruption. 
{The Br-$\gamma$ line recorded at our highest 
resolution of $R\sim3000$ clearly shows a double-peaked profile at 
one epoch, with a separation of $\sim500$\vunit.}
This is consistent with the signature
for emission lines originating from an AD 
\citep{horne}. Since the dip between the peaks is not very deep, 
a system at low inclination ($\ltsimeq30^\circ$) is indicated \citep{horne}
{if the emission comes from the AD only.}

Velocities of up to 2,000\vunit\ are also indicated by the 
\pion{He}{i} 2.0587\mic\ line, although there are dramatic
changes in the line profile. (Based on the close 
agreements of the central wavelengths of the centers of the \pion{H}{i} 
lines with their laboratory wavelengths, we conclude that the radial 
velocity of center of mass of the \vcen\ system is no more 
than several tens of \vunit). The July spectrum of this line 
displays a complex profile, with absorption extending from $\sim0$ to
$-2000$\vunit, sharp absorption 
{troughs} at  $-300$ and $-1,400$\vunit, 
and weak emission extending from $\sim0$ to at least 1600\vunit. 
{The troughs are replicated in the \pion{He}{i} 
$\lambda2.1138$\mic\ profile}  (see Fig.~\ref{vel}).
In the August spectrum only a single absorption 
{trough} is present, 
at $-1,300$\vunit, the wing to higher negative velocities has weakened, 
and the emission now is much more extensive, from $-800$ to $+1600$\vunit.

The \pion{He}{i} triplet at 1.0833\mic\ is almost entirely in emission, 
but is more difficult to characterize quantitatively, due to the 
lower resolution at which it was observed and its blending on the long 
wavelength side with  Pa-$\gamma$ (1.0941\mic). In the July~1 
spectrum only a single weak P~Cygni absorption is present at $-1,600$\vunit,
close in velocity to the more blueshifted absorption in the \pion{He}{i}
singlet profile, and with an absorption wing extending to $\sim-2,000$\vunit,
similar to the singlet. In the August~2 spectrum absorption is completely 
absent, and the emission extends to $\sim-2,000$\vunit. At both times 
the FWHM of the line emission is $\sim1,200$\vunit. 
In contrast, \cite{aydib} found relatively narrow ($\mbox{FWHM} < 300$\vunit)
\pion{He}{i} emission lines in 
an optical spectrum obtained on July 19.86~UT.
We note that the  (deconvolved) FWHM of the \pion{O}{i} line at 
1.1289\mic, present in the August 2 spectrum, is $\sim700$\vunit.

The cause of the P Cyg profile in the \pion{He}{i} 2.0587\mic\ line 
observed on July 1 is puzzling,
{but its reality is beyond doubt
(see Fig.~\ref{vel}).} We do not find a similar profile in 
the literature, though NIR spectra of DNe during outburst are rare 
\citep[e.g.][]{howell}. In quiescence such P Cyg profiles are not 
generally seen \citep[see e.g. the DN sample observed by][]{dhillon}. 
It is possible that the 2019 eruption has generated 
two separate outflows, as witnessed in the July~1 profile of the 
singlet line.  However radiative transfer associated with this line 
is complex \citep[see][]{geballe}, and what we have observed may 
be a consequence of it rather than multiple outflows.

Many ultraviolet resonance lines
(\pion{C}{iv} 1549\AA, \pion{Si}{iv} 1397\AA, \pion{N}{v} 1240\AA)
in DNe often display P~Cyg profiles, implying the presence of
mass-loss in stellar winds \citep[e.g.][and references therein]{holm,warner}.
\pion{He}{ii} 1640\AA\ and \pion{N}{iv} 1719\AA\ have also been found to
have P~Cyg features, in RW~Sex \citep{drew}, and it is possible that the
\pion{He}{ii} recombination lines (1640\AA, 4686\AA) are produced
in both the disk and the wind \citep{warner}. We may thus speculate
about the presence of a wind in this system which is responsible
for the P~Cyg profile in \pion{He}{i} 2.0587\mic.

\begin{figure}
\includegraphics[width=7.cm,keepaspectratio]{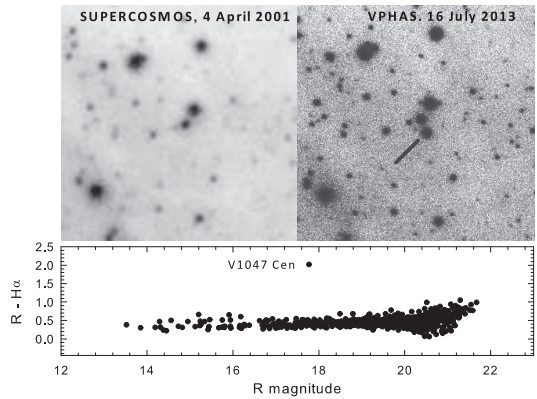}
\caption{$75''$ (approx) square SUPERCOSMOS (top left) and VPHAS 
(top right) H$\alpha$ images of the \vcen\ field. 
North is up, east is right.
The nova is arrowed on the VPHAS image.
Bottom: ($R-\mbox{H}\alpha)-R$ diagram for \vcen\ field, showing
large H$\alpha$ excess in \vcen. 
\label{cosmos}}
\end{figure}

\vcen\ also shows an intriguing H$\alpha$ excess in images 
obtained prior to its current rebrightening.
A SUPERCOSMOS \citep{hambly} H$\alpha$ image of the \vcen\
field, obtained in 2001, shows no source at the position of the nova
(see Fig.~\ref{cosmos}). On the other hand a VPHAS \citep{vphas} image
shows that \vcen\ had a strong H$\alpha$ excess in 2013, 
several years after its 2005 eruption,
{but before the 2019 event.
The H$\alpha$ excess in \vcen\ is comparable with, or even stronger 
than, the excess in novae with giant secondaries and which show strong
H$\alpha$ in the SUPERCOSMOS images, as well as very strong H$\alpha$ 
emission in their quiescent spectra \citep{anupama}.}
The presence of an excess
may lend support to our DN interpretation, since not only DNe in quiescence, 
but also old CNe and their remnants, show H$\alpha$ emission in their 
spectra \citep{stauffer,shafter}. 
We also note that \pion{N}{ii} (6548\AA, 6584\AA), known 
to be seen in CNe,
could also be contributing partially to the H$\alpha$ images.
Nonetheless the strength of the H$\alpha$ excess is puzzling and 
its evolution should be monitored.

There is some evidence for the
presence of the \fion{Si}{vi} 1.9641\mic\
coronal line in the earlier spectrum (see Fig.~\ref{gemini}).
We note that a number of the \pion{H}{i} recombination lines have
weak features on their red wings from \pion{He}{ii} but this
is not the case for Br-$\delta$.
The ionization potential of \pion{Si}{v} is 167~eV, so the production of
\pion{Si}{vi} requires high energy conditions such as might occur in a shock.
We have noted the likely presence of winds having different velocities;
conditions for coronal emission might well occur in regions where the winds
collide. We note that \fion{Si}{vi} 1.9641\mic\ emission has previously
been detected in the CV TT~Ari \citep{ramseyer}.

\section{Discussion}
\label{disc}

If the 2019 April event in \vcen\ is a DN
eruption in a CN system, then we are witnessing
a thermal instability in the AD. In their model of the disk-instability-induced 
DN eruptions in GK~Per, \cite{kim} found
that AD temperatures in the range 
6000--10000~K are expected. In order for the IR emission to
reflect such temperatures, 
dereddening by $E(B-V)\gtsimeq1.6$ is required, 
for both the July and August spectra. This implies either that the 
reddening (and hence distance) to \vcen\ is at the higher end of the range
implied by the {\it Gaia} data, or that what we are seeing is not a
thermal instability event.

Only seven CNe are known to
have exhibited DN outbursts within a few decades of their
original CN outburst \citep[see][]{vogt,honeycutt}. The best
documented DN outbursts are those of GK~Per (CN 1901),
V1017~Sgr (CN 1919) and V446~Her (CN 1960), while their subsequent
DN eruptions were first recorded decades later: GK Per in 1948
\citep{bianchini}, V446~Her \citep[around 1991;][]{honeycutt}
and V1017~Sgr \citep[1973;][]{warner}. While DN outbursts have been
observed regularly in GK~Per and V446~Her \citep{bianchini, honeycutt},
only two DN outbursts are known to have occurred 
for V1017~Sgr, though it displayed a DN eruption in 1901 prior to its
CN outburst \citep{salazar}. The durations  of individual DN
events in these CN systems also vary substantially, lasting typically
about 2~months for GK~Per \citep{bianchini}, 10--20 days for V446~Her
\citep{honeycutt} and $\sim200$ days for V1017~Sgr.

In this context the 2019 DN outburst in V1047~Cen has two notable
characteristics. The duration of the outburst is relatively
long, and as of mid-October 2019 the decline phase has yet to commence.
Further, the DN event occurred only 14~years after the CN eruption,
the shortest recorded gap between CN and DN eruptions.
Explaining this behavior within the framework of the
``hibernation'' paradigm \citep{shara-apj86} may be challenging.
In that theory for cyclic nova evolution, the mass
accretion rate $\dot{M}$ remains high
$(\dot{M}\sim10^{-8}$\Msun~yr$^{-1})$ on a time-scale of a few
centuries after the CN eruption due to the irradiation of
the secondary star by the still-hot WD
\citep{shara-apj86,kovetz}. Eventually $\dot{M}$ declines,
taking the AD through
the DN phase and into hibernation, from which the CV reawakens
after several millennia for the next CN eruption. The theory is
supported by the detection of ancient CN remnants around the
DNe Z~Cam \citep{shara-nat}, AT~Cnc \citep{shara-apj12}, and
also by DNe outbursts preceding the main CN eruption of
Nova Cen 2009 \citep{mroz}. A DN eruption is not expected to
occur during the high $\dot{M}$ phase that lasts for a few
centuries after the CN outburst, as DN outbursts are believed
to result from thermal instabilities associated with hydrogen
ionization in the AD occurring during a regime of {\em low}
mass accretion rate
\citep[$\dot{M}\sim10^{-10}$\Msun~yr$^{-1}$;][]{osaki,cannizzo}.
Thus a special set of circumstances is needed to account for the early DN
outburst in \vcen. In the case of GK~Per, DN eruptions are not
surprising \citep{schreiber} as it has a large AD \citep{kim},
consistent with its 2-day orbital period. Thus the critical mass
transfer rate $\dot{M}_{\rm crit}$ below which outbursts are possible
is large \citep[note that $\dot{M}_{\rm crit}$ scales as
$(R_{10})^{2.6}$, where $R_{10}$ is the outer disk
radius in units of $10^{10}$~cm;][]{cannizzo}. Thus DN
outbursts are possible in GK~Per even at high accretion rates
\citep{kim}. It is possible that similar circumstances are responsible
for the DN outburst of \vcen\ but a robust determination of
the orbital period is needed to verify this.

In summary, we have presented NIR spectra of the 2019 outburst of
\vcen. The presence of mostly H and \pion{He}{i} lines in the spectra,
the presence of double-peaked line profiles, and the nature of the
light curve all suggest that a DN outburst has occurred.
Such an event is rare, especially coming so soon after the CN
outburst, and challenging to explain. The present and future outbursts
of \vcen\ could provide an opportunity to probe the CN to DN
transition phenomenon and also test the generic mechanisms responsible
for DN outbursts.

\acknowledgments

This paper is largely based on observations obtained for program
GS-2019A-DD-111 of the Gemini Observatory, which is operated by the
Association of Universities for Research in Astronomy, Inc., under
a cooperative agreement with the NSF on behalf
of the Gemini partnership: the National Science Foundation (United States),
National Research Council (Canada), CONICYT (Chile), Ministerio de Ciencia,
Tecnolog\'{i}a e Innovaci\'{o}n Productiva (Argentina), Minist\'{e}rio da
Ci\^{e}ncia, Tecnologia e Inova\c{c}\~{a}o (Brazil), and Korea Astronomy
and Space Science Institute (Republic of Korea).

The OGLE project has received funding from the National Science Centre, Poland,
grant MAESTRO 2014/14/A/ST9/00121 to AU.

This work has made use of data from the European Space Agency (ESA) mission {\it Gaia} 
processed by the {\it Gaia} Data Processing and Analysis Consortium (DPAC),
funding for which has been provided by national institutions, in particular the institutions
participating in the {\it Gaia} Multilateral Agreement.

Based on data products from observations made with ESO Telescopes at the
La Silla Paranal Observatory under programme ID~177.D-3023, as part of
the VST Photometric H$\alpha$ Survey of the Southern Galactic Plane and
Bulge (VPHAS+, www.vphas.eu).

Nova research at Stony Brook is supported by a grant from
the NSF. FMW acknowledges support from Stony Brook University for continued
participation in the SMARTS consortium.
DPKB is supported by a CSIR Emeritus Scientist grant-in-aid hosted
by the Physical Research Laboratory, Ahmedabad.
CEW acknowledges partial support from a USRA/NASA SOFIA contract.
RDG is supported by NASA and the United States Air Force.
KLP acknowledges support from the UK Space Agency.
SS acknowledges support from a NASA Theory grant to ASU.
DAHB thanks the National Research Foundation for research support.

We acknowledge with thanks the variable star observations from the AAVSO
International Database contributed by observers worldwide and used in this research.

\facilities{Gemini South (FLAMINGOS-2), OGLE, Gaia, ESO, SMARTS, AAVSO}

\software{IRAF \citep{iraf1,iraf2}; Figaro  \citep{figaro}}

\end{document}